\documentclass[groupedaddress,superscriptaddress,floatfix,10pt,article,onecolumn,notitlepage]{revtex4-2}
\usepackage[english]{babel}
\usepackage[utf8]{inputenc}
\usepackage[T1]{fontenc}
\usepackage{lmodern}

\usepackage{amsmath}
\usepackage{amssymb}
\usepackage{mathptmx}
\usepackage{helvet}
\usepackage{graphicx}
\usepackage{color}
\usepackage{psfrag}
\usepackage{bm}
\usepackage{xcolor}

\usepackage{bm}
\usepackage{braket}

\usepackage{caption}
\usepackage{subcaption}

\usepackage[a4paper,margin=2.5cm]{geometry}

\usepackage{ulem}
\usepackage{hyperref}

\long\def\symbolfootnote[#1]#2{\begingroup%
\def\thefootnote{\fnsymbol{footnote}}\footnote[#1]{#2}\endgroup} 
\def\blfootnote{\xdef\@thefnmark{}\@footnotetext}

\definecolor{purp}{rgb}{0.5,0,0.5}
\definecolor{darkgreen}{rgb}{0.1,0.7,0}
\definecolor{orange}{rgb}{1,0.6,0}
\newcommand{\bl}[1]{{\textcolor{black}{#1}}} %% FIX

\newcommand{\be}{\begin{equation}}
\newcommand{\ee}{\end{equation}}
\newcommand{\bi}{\begin{itemize}}
\newcommand{\ei}{\end{itemize}}
\newcommand{\bea}{\begin{eqnarray}}
\newcommand{\eea}{\end{eqnarray}}

\begin{document}
\captionsetup[figure]{labelfont={bf},labelformat={default},name={Fig.},justification=raggedright,singlelinecheck=false,format=hang}
\title{Emergence of nonclassical radiation in strongly laser-driven quantum systems} 

\author{Ivan Gonoskov}
\email[]{ivan.gonoskov@uni-jena.de}
\affiliation{Institute of Physical Chemistry, Friedrich Schiller University Jena, Lessingstr. 4, 07743 Jena, Germany}

\author{Christian H\"unecke}
%\email[]{christian.huenecke@uni-jena.de}
\affiliation{Institute of Physical Chemistry, Friedrich Schiller University Jena, Lessingstr. 4, 07743 Jena, Germany}

\author{Stefanie Gräfe}
\email{s.graefe@uni-jena.de}
\affiliation{Institute of Physical Chemistry, Friedrich Schiller University Jena, Lessingstr. 4, 07743 Jena, Germany}
\affiliation{Fraunhofer Institute for Applied Optics and Precision Engineering, Albert-Einstein-Str. 7, 07745 Jena, Germany}

\date{\today}

\begin{abstract}
\textbf{
\bl{We present an analytical framework for the emergence of nonclassical radiation in strongly laser-driven quantum systems, with a focus on high-order harmonic generation (HHG). Starting from a Pauli–Fierz description, we employ a parametric factorization of the coupled light–matter wavefunction that reduces the dynamics to coupled equations for a field-driven electronic state and a quantized light mode. Within this framework, we identify a simple and predictive mechanism for nonclassicality: it originates from the nonlinear dependence of the electronic dipole response on the light-mode coordinate. An approximately constant dipole yields coherent radiation, a linear dependence produces squeezing, and higher-order nonlinearities give rise to Wigner-function negativity. We illustrate this mechanism for atomic and molecular model systems and analyze its scaling in multi-emitter configurations, indicating routes toward high-photon-number nonclassical radiation in HHG. Our results provide a transparent connection between strong-field dynamics and quantum-optical properties of emitted light, offering a basis for engineering nonclassical states in intense laser–matter interactions.}	
}
\end{abstract}
\maketitle
%\section{}
%\pacs{}

\bl{Nonclassical states of light are a key resource for quantum technologies, yet existing platforms typically offer limited tunability and low photon numbers \cite{Kl-Rev}. In contrast, strong-field processes such as high-order harmonic generation (HHG) provide bright, broadband radiation spanning from the infrared to the extreme ultraviolet. High harmonics do not only constitute a bright tunable table-top source of coherent high-frequency radiation, but the broad, coherent spectrum, spanning many octaves, also is the way to obtain attosecond pulses \cite{Atto1}. Recent experiments have revealed quantum-optical signatures in HHG \cite{Gonoskov, Tzallas}, including photon-number correlations, squeezing, and entanglement \cite{Theidel, ParisBN}, raising the question of how nonclassicality emerges in this highly nonlinear, many-photon regime.}

\bl{Theoretical descriptions of HHG in the quantum-optical domain have advanced significantly in recent years \cite{Gonoskov, Rivera, Lamprou, Lewenstein, Stammer_PRL, Stammer0, Stammer_PRX, Gonoskov_2021, Ivanov, Dan, Gorlach2, Rubio-Neufeld}. Approaches based on time-dependent Schrödinger equations, perturbative expansions, or quantum-state engineering protocols have demonstrated the presence of entanglement and nonclassical features in both the driving and emitted fields. However, these descriptions often rely on conditioning, specific input states, or perturbative treatments, and a transparent mechanism linking the strong-field electron dynamics to the quantum properties of the emitted radiation remains less clear.}

\bl{In this work, we present an analytical framework that directly connects the electron dynamics in a strong laser field to the quantum state of the emitted radiation. Key idea is a parametric factorization of the coupled system of light and matter into a \textit{field-driven} electronic state and the light state \cite{Gonoskov_2021, Gonoskov_2024}, which incorporates the system's (driven) dynamics as a finite perturbation to the light field. In contrast to other works, our ansatz is directly based on the time-dependent Schr\"odinger equation (TDSE), without the need of conditioning or post-selection techniques. %Building on a parametric factorization of the light–matter wavefunction, we reduce the coupled problem to a set of equations for a field-driven electronic state and a quantized light mode. 
While parametric factorization has been introduced previously \cite{Gonoskov_2021, Ivanov}, its physical implications for the emergence of nonclassical radiation have not been identified.  Here we show that the nonlinear dependence of the dipole moment on the light-mode coordinate provides a direct and predictive criterion for nonclassicality: %Within this framework, we identify a simple physical mechanism governing nonclassicality: it is controlled by the dependence of the electronic dipole response on the coordinate of the quantized light mode.
%We show that this dependence provides a direct criterion for the nature of the emitted radiation. 
If the dipole response is approximately independent of the light-mode coordinate, the emitted radiation remains close to a coherent state. A linear dependence leads to squeezing, while higher-order nonlinearities generate strongly nonclassical states characterized by Wigner-function negativity. Consequently, nonclassicality in HHG emerges from the nonlinear dependence of the system's dipole moment on the light mode.}

\bl{The parametric factorization reduces the original problem to coupled 1D equations, enabling efficient numerical implementation (see Appendix). Compared to direct TDSE simulations, this approach significantly reduces computational complexity and allows treatment of multi-emitter systems. We demonstrate this mechanism for atomic and molecular model systems and analyze its behavior in multi-emitter configurations, highlighting the conditions under which nonclassical features are enhanced. Our results provide a transparent connection between strong-field electron dynamics and quantum-optical properties of HHG radiation. This establishes a physically intuitive framework for understanding and controlling nonclassical light generation in intense laser–matter interactions.}

Starting point of the analytical model is the Pauli-Fierz (PF) Hamiltonian, providing a fully quantum description of nonrelativistically charged particles coupled with electromagnetic fields.
\begin{equation}\label{eq:Pauli_Fierz}
	\hat{H}=\sum\limits_{k}\frac{1}{2m_{k}}\left(\,\hat{\vec{p}}_{k}+\frac{e_{k}}{c}\sum\limits_{j}\hat{\vec{A}}_{j}\right)^{2}+\sum\limits_{k\,k'}U_{k\,k'}+\sum\limits_{j}\hbar \omega_{j}\hat{N}_{j}\;,
\end{equation}

In this equation, $e_{k}$ and $m_{k}$ are the charges and masses of the particles (index $k$), with $U_{k\,k'}$ containing both, mutual particle-particle interaction and interaction of each particle with external longitudinal fields. The quantized electromagnetic field modes $j$ enter the Hamiltonian via $\omega_j \hat{N}_{j}$ ($\omega_j$ is the frequency and $\hat{N}_j$ the photon number operator). In intense fields, $\langle \hat{H}_{el} \rangle  \sim \langle \hat{W}_{int}\rangle \ll \langle \hat{H}_{field}\rangle $, and we assume that the quantum state of light changes only weakly during the interaction compared to the initial state of light (the interacting electron is a small perturbation to the strong field) \cite{Tzur_2023}. This allows us to describe the total wavefunction containing both, the system’s ($x$) and light’s ($q$) degrees of freedom in a product form:
 
\begin{equation}\label{eq:factorization}
\Psi(x,q_j,t)=\psi^{el} (x,\beta_j q_j,t)\cdot \phi^{light} (q_j,\beta_j,t) + \mathcal{O}(\beta_j),
\end{equation}	
			
where $\beta_j=c\sqrt{(2\pi\hbar/\omega_j V)}$ and $V$ is the quantization volume. While commonly the wavefunction is factorized into a purely electronic wavefunction and function for the light’s state, in above equation, $\psi^{el} (x, \beta_j q_j, t)$ is the field-dressed electronic wavefunction of the system. 

This ansatz is very beneficial from the strong-field physics’ perspective, as it includes implicitly the impact of the strong fundamental driving field on the system and allows us to describe the systems’ response to the light field as a finite perturbation to the field (as $\beta_j q_j$ is proportional to te amplitude of the $j-$th mode). 
Specifically, field’s vector potential can be divided into a predominantly classical part, $A_{class}(t)$, the driving laser, and the modes of the emitted (harmonic) light, $\hat{A}_{\Omega}$. For the primary derivation, we consider one mode of the emitted light $\Omega$ with coordinate $q$, \bl{parameter $\beta$}, and restrict ourselves to fields linearly polarized along the $x$-direction. \bl{We emphasize that we do not need to fundamentally restrict the derivation for the driving field to be purely classical. So far, it has remained unclear whether the nonclassical character of the harmonics (as experimentally observed) originates solely from “fluctuations” or nonclassical properties of the driver, which are “enhanced” upon upconversion \cite{Tzur_2023, Gorlach, Fang, Tzur_2024}. As we will demonstrate here, even a classical driver $A_{class}(t)$ can lead to generation of harmonics with nonclassical properties (squeezing, Wigner-negativity).}
 We further apply a quantum optical analogue of the gauge transformation $\exp\{\mathrm{i}x\, e/(c\hbar) [A_{class}(t)+ \hat{A}_{\Omega}(t)]\}$ to write the time-dependent Schr\"odinger equation within the dipole approximation as
\begin{equation}  
\mathrm{i}\hbar \frac{\partial}{\partial t} \Psi(x,q,t)=\left[\hat{H}_0+ e x\,[F_{class}(t)+ \hat{F}_{\Omega}(t)]\right] \Psi(x,q,t),	
\end{equation}			
where $\hat{H}_0$ is the initial unperturbed Hamiltonian, $F_{classs}(t)=-(1/c)\cdot \dot{A}_{class} (t)$ and the force operator 
\begin{equation}
	\label{eq:force_operator}
\hat{F}_{\Omega}=  \beta q \frac{\Omega}{c} \sin \Omega t +i \beta \frac{\Omega}{c}\frac{\partial}{\partial q} \cos \Omega t	
\end{equation}				
which is linear with respect to $q$ and $\partial/\partial q$. \bl{The advantage of the parametric factorization is that $\hat{F}_{\Omega}(t)\,\Psi(x,q,t) \approx{}\psi^{el} (x,\beta q,t)\cdot{}\hat{F}_{\Omega}(t)\,\phi^{light}(q,\beta ,t)$ neglecting small perturbations of higher order (see also appendix)}. Consequently, we can decouple the total Schr\"odinger equation for the combined system of light and matter into two equations, one for the field-driven system
\begin{equation}\label{eq:TDSE-x}
\mathrm{i}\hbar \frac{\partial}{\partial t} \psi^{el} (x,\beta q,t)= \hat{H}_0 \psi^{el}(x,\beta q,t)+ e x \left[F_{class}(t)+F_{\phi^{light}} (\beta q,t)]\right] \psi^{el}(x,\beta q,t)	
\end{equation}
and one for the light mode, 
\begin{equation}\label{eq:TDSE-q}
\mathrm{i}\hbar \frac{\partial}{\partial t} \phi^{light}(q,\beta ,t)= + e \langle x(\beta q,t)\rangle_{\psi^{el}}  \left[F_{class}(t) + \hat{F}_{\Omega}(t) \right] \phi^{light}(q,\beta,t) 		
\end{equation}

These equations are parametrically connected by the electron coordinate expectation value  
\begin{equation}\label{eq:dipole}
\langle x(\beta q,t)\rangle_{\psi^{el}} =\int x |\psi^{el}(x,\beta q,t)|^2 dx	.			
\end{equation}

The parametric-connection approximation introduced in \cite{Gonoskov_2021} corresponds to an advanced quantum optical analogue of the Born-Oppenheimer approximation for electron-nuclear dynamics following from the exact factorization introduced by the group of E.K.U. Gross in \cite{EF1}). Entering the system’s time-dependent Schrödinger equation \ref{eq:TDSE-x} is the (unitary-transformed) force operator (local in $q$)
\begin{equation}
F_{\phi^{light}} (\beta q,t)= \left(\phi^{light}(q,\beta ,t)\right)^{-1} \, \hat{F}_{\Omega}(t) \, \phi^{light}(q,\beta ,t) 	
\end{equation}
(in a time interval, where $\phi^{light}\neq 0$); \bl{for nearly coherent states, the unitary-transformed operator in our case is proportional to $ \frac{\Omega}{c} \beta q \sin \Omega t$.} The detailed derivation can be found in the appendix. This system transforms the original 2-dimensional time-dependent Schr\"odinger equation into a system of two 1-dimensional equations connected parametrically via the $\beta{}q$-dependence in {$\psi^{el} \left( x,\,{\beta{}q}\,,t\right)$. The derivation is valid for relatively small intensities of emitted quantized modes and small/moderate changes in field states $\phi^{light}$ compared to a Gaussian (coherent-state) distribution. However, it can be further improved and generalized in various ways, for example to take into account the collective effects of multiple atomic/molecular systems and multiple resonator modes. The separation ansatz provides the potential for significant acceleration of the numerical calculations compared to the typical numerical solution of the original time-dependent Schr\"odinger equation, with the numerical schemes being described in the appendix.

Thus, in this system of equations, the nonlinear dependence on $q$ arises from the $q$-dependence of the oscillating dipole moment of an electron in state $\psi^{el} (x,\beta q,t)$. Consequently, the nonlinear terms in eq.~\ref{eq:dipole} can lead to the generation of non-classical states of light, even if the initial state is a coherent or a vacuum state. To demonstrate this, for visualization we expand the time-dependent dipole moment into a series 
\begin{equation}\label{eq:dipole-series}
	\langle x(\beta q, t)\rangle _{\psi^{el}} \approx \langle x(t) \rangle_{\psi} \,|_{F_{class}(t)} + (\beta{}q)\cdot{}f_{1}(t) + (\beta{}q)^{2}\cdot{}f_{2}(t) + \ldots\;,
\end{equation}
where $f_{1}(t)\sim\cos \left( \Omega t\right)\cdot\dfrac{\partial \langle x\rangle_{\psi^{el}} }{\partial F}|_{F=F_{class}(t)}$ (and higher order terms correspondingly, see appendix). Figure \ref{fig:overview} illustrates the central mechanism %provides an overview 
of the emergence of nonclassical light of a strongly driven Ca-atom model: clearly, constant, almost $q$-independent dipole moments generate harmonic radiation corresponding to an almost coherent state with high photon number (panel A). With a slight linear $q$-dependence of the dipole moment, squeezed light states with high photon number can be generated (panel B). Strikingly, for dipole moments with a nonlinear dependence on $q$ (\bl{in panel C, the dependence is approximated to within cubic terms}), nonclassical light can be generated, as can be seen from the pronounced negative regions of the corresponding Wigner function. Thus, the nonlinearity in $q$ of the oscillating dipole moment is a necessary (but not sufficient) condition for nonclassical light generation. \bl{We note that the choice of laser and system parameters takes the following conditions into account: (i) the total ionization for longer driving wavelengths is lower; (ii) in previous work on HHG in semiconductors \cite{Gonoskov_2024}, we found the nonclassical character of the emitted harmonics to increase for longer driving wavelengths; (iii) for longer driving wavelengths, several higher harmonics lie in the visible range, where highly accurate detectors are available; (iv) in agreement with the recent results from Ivanov and coworkers \cite{Ivanov}, we find that the nonclassical character of the emitted radiation is enhanced when hitting a resonance (excited electronic state of the matter). For the (model) systems considered here, these criteria can be met best by the parameters specified in the figure caption.}

\begin{figure}
	\includegraphics[width=\columnwidth]{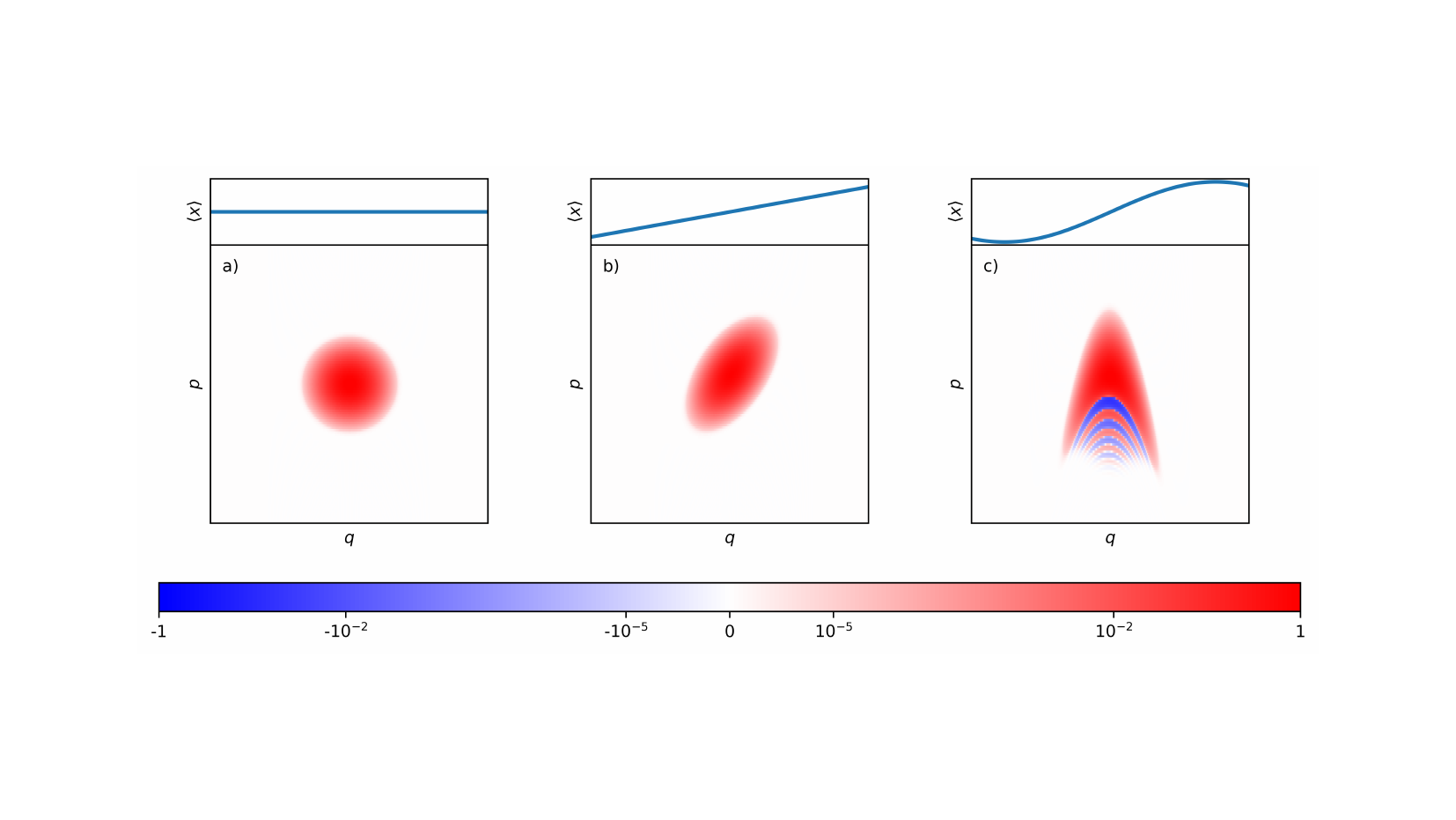}
	\caption{Visualization of the emergence of nonclassicality in high-order harmonic generation to result from the nonlinear $q$- dependence of the oscillating dipole moment. The upper row displays the dipole moments expanded into a series (truncated after the terms as labeled) (eq.~\ref{eq:dipole-series}) while the lower row shows the resulting output state of light in phase space, represented by its Wigner function. (a) Constant, almost $q$-independent dipole moment leads to (below): emitted light represented by a Wigner function of an almost coherent state with high photon number (here $\Omega=5$th harmonic). (b) Linear $q$-dependence of dipole moment can lead to Wigner function of a squeezed state with high photon number ($\Omega=5$th harmonic). (c) Dipole moment containing terms nonlinear in $q$, here for an approximately \bl{cubic} dependence on $q$, can lead to the generation of nonclassical light, as represented by a Wigner function with negative values/regions ($\Omega=5$th harmonic).  Initial conditions: product state of light and matter, with the initial system being in the second excited state for the H-atom; Laser parameters: $F_{class}(t)\;$ ($\lambda$ = 4000 nm, T$_{p}$ = 4 cycles,  $I = 4\cdot 10^{12}$ W/cm$^{2}$). The quantization parameter is $\beta = 0.274 \text{au.}$. The average number of photons in the output state is approximately $N_{\Omega}\approx{\bar{q}^{2}/2}$}\label{fig:overview}
\end{figure}

With the theoretical description established, we next analyze under which conditions in the field-driven electronic system a nonlinear dependence in $q$ of the dipole moment results. Similar to the work by Ivanov and coworkers \cite{Ivanov}, we find that light-matter entanglement, and thus, generation of nonclassical light of frequency $\Omega$, is favorable when this corresponds to resonant transitions in the system, i.e., $\Omega \approx \Delta E/\hbar$. Thereby, the population and phase of the ground and excited electronic states involved determine the characteristic nonclassical properties of the generated harmonics, as shown in Fig.~\ref{fig:Wigner-resonances}: the panels (a) show the resulting Wigner functions of the Ca model atom initially in the electronic ground state (a1) and being in a superposition state of an equally populated ground and excited electronic state (a2) ; panels (b1) and (b2) show the same for a diatomic model molecule with fixed internuclear distance. Ultimately, panels (c1, c2) display the corresponding Wigner function for the $\Omega=13$th harmonic generated from the laser-driven hydrogen atom, demonstrating nonclassical light in the UV region ($\lambda_{13}=171$ nm). As the output non-classical states strongly depend on the populations and phases of the ground/excited initial states in the system under consideration, this servers as an additional control determining the properties of the generated nonclassical light.

\begin{figure}
	\includegraphics[width=0.8\columnwidth]{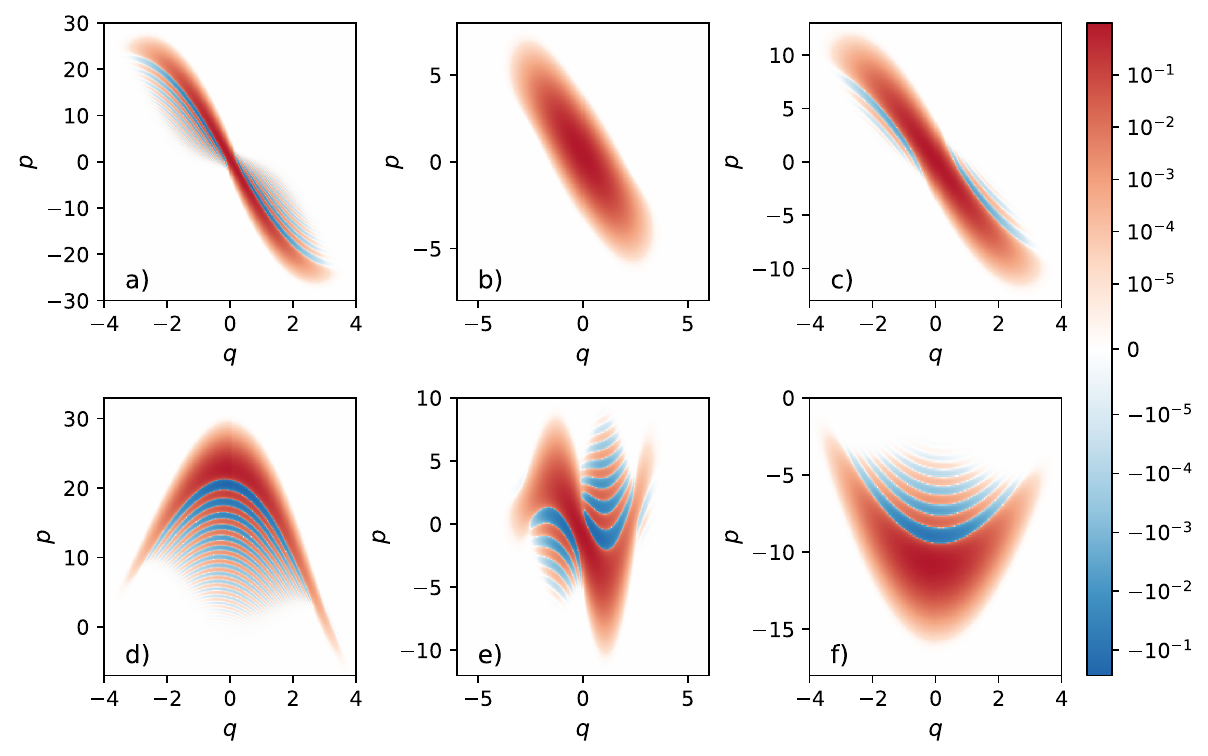} 
	\caption{{Gallery of Wigner functions after the interaction with a laser pulse for different systems and initial conditions. (a, d) atomic system mimicking the Ca atom; (b, e) molecular system with fixed internuclear distance; (c, f) hydrogen atom-like potential. The upper row shows the results when starting with the electronic ground state of the driven system while the lower row the corresponding results when starting with a superposition state: (a) Atom initially in the electronic ground state; (d) initial atomic state corresponding to a superposition state of the ground state and excited state. The following parameters are employed:  $\Omega= 7; F_{class}(t):\;$ $\lambda$ = 3750 nm, T$_{p}$ = 2 cycles,  $I = 2\cdot 10^{12}$ W/cm$^{2}$ and $\beta = 0.55 \text{au.}$}; (b, e) Same for the corresponding molecular model system, with $\Omega= 3; F_{class}(t):\;$ $\lambda$ = 3750 nm, T$_{p}$ = 2 cycles,  $I = 1\cdot 10^{12}$ W/cm$^{2}$ and $\beta = 0.82$ a.u.; (c, f) for the H-atom model with $\Omega= 13; F_{class}(t):\;$ $\lambda$ = 2227 nm, T$_{p}$ = 2 cycles,  $I = 1\cdot 10^{13}$ W/cm$^{2}$ and $\beta = 0.41 \text{au.}$  }\label{fig:Wigner-resonances}
\end{figure}

So far, we have considered solely the interaction of single emitters/matter systems with intense light. An additional advantage of the presented theory is its universal scalability for an arbitrary number of quantum systems and quantized light modes. This case naturally includes mode-to-mode entanglement. Let us consider exemplary an ensemble of spatially isolated non-Coulomb-interacting emitters in a resonator \cite{LargeNum}. Each quantum system $n$ is strongly driven by the intense laser field and generates a certain harmonic of frequency $\Omega$ via its oscillating dipole $d_{n} (\beta q,t )=e_j \langle x(\beta q, t)\rangle_j$. Such a superposition $D(\beta{q},t)=\sum_n {d_{n}(\beta{q},t)}$ can, under certain conditions, lead to the generation of intense non-classical light, as we demonstrate in Fig.~\ref{fig:intense}. \bl{Of course, such a scenario constitutes an idealized assumption, however, is evoked to demonstrate the scalability of our approach.} As explained above, also the multi-dipole systems feature resonances and, therefore, lead to relatively high dipole nonlinearity. Due to the large number of emitters under consideration, nonclassicality is noticeable even for intense output states of light. \bl{We emphasize once again that we do not need to evoke the assumption that the driver or its harmonics are initially coherent, vacuum (Gaussian), or classical fields.} For nonclassical initial harmonic states instead of the usual coherent/vacuum states considered so far, the nonclassicality of the output radiation can be significantly enhanced (see Fig.~\ref{fig:intense} b), paving the way to generating bright, highly nonclassical light. 

\begin{figure}
	\includegraphics[width=0.85\columnwidth]{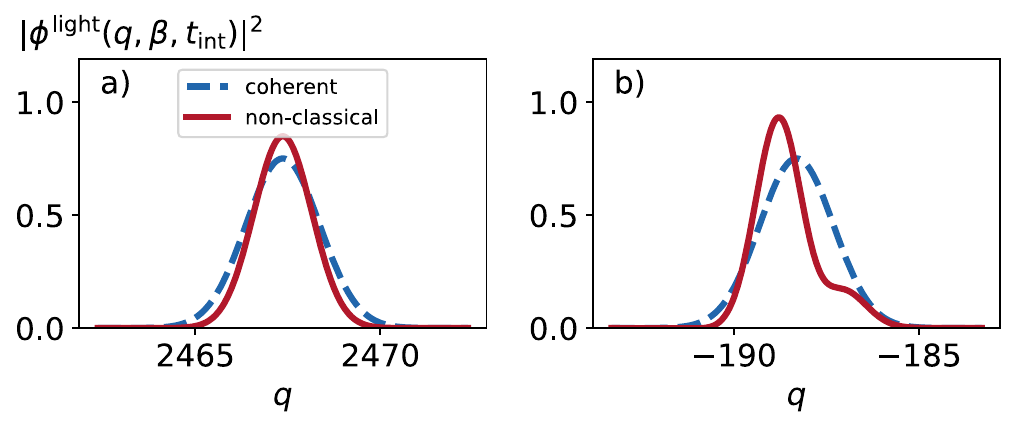}
	\caption{Nonclassical output states for $\Omega=5$th harmonic generated from the multi-emitter case with $N_e$ spatially isolated (non-Coulomb interacting) emitters. Each emitter is a 1D model CaO model molecule).  The parameters of the driving laser pulse are $F_{class}=3.4\cdot{}10^{-2}\,\text{(a.u.)}$, $I = 4\cdot 10^{13}$ W/cm$^{2}$, $\lambda=1945\,\text{nm}$. The solid line corresponds to the solution based on the parametric factorization. The dashed line shows the corresponding coherent state for comparison (no $\beta{q}$-dependence in the dipole). (a) The initial state is a vacuum (Gaussian) state and $\beta=10^{-8}$ considering $N_e=10^{11}$ emitters. (b) Initial state consisting of a weakly nonclassical state, represented by a vacuum (Gaussian) with several percent of 1-photon (7\%), 2-photon (3\%), and 3-photon (1.7\%) states, with $\beta=10^{-7}$ and $N_e=10^{9}$ emitters. 
	The calculations are performed by fitting the dipole up to the 5-th order (eq.~\ref{eq:dipole-series})). As the average number of photons in the output mode is $N_{\Omega}\approx{\bar{q}^{2}/2}$, it can be seen that very bright nonclassical states can be obtained.} \label{fig:intense}
\end{figure}

In summary, we have presented a rigorous model describing the generation of nonclassical electromagnetic radiation in strongly driven systems. It is based on the factorization and subsequent separation of the time-dependent Schr\"odinger equation for light and matter into a system of equations, with one equation for the field-driven system and equation(s) for the quantized light mode(s), parametrically coupled via the oscillating dipole of the driven system $\langle x(\beta q,t)\rangle_{\psi^{el}}$. With this theory, we provided a mechanism for the emergence of nonclassical (high-harmonic) light generation in terms of a nonlinear dependence of this dipole moment $d(\beta q,t)$ on the light mode's coordinate $q$, which becomes particularly pronounced, when the harmonic frequency $\Omega$ corresponds to an electronic resonance in the system. The parametric factorization reduces the original problem to coupled 1D equations, enabling efficient numerical implementation as this reduces computational complexity significantly. Thus, the herein introduced parametric factorization provides the possibility to numerically treat several light modes or macroscopically large multi-dipole systems, which we showed to produce for certain conditions highly intense and significantly non-classical radiation. Moreover, as the theory naturally includes entanglement of light and matter, it provides a natural and universal explanation on the origin of the nonclassical nature of radiation in driven systems, \bl{allowing to be extended to further systems,} including atoms, molecules, and possibly semiconductors or even exotic materials/medium such as quantum materials.

\section*{Acknowledgments}

The authors highly acknowledge funding from the German Research Foundation (DFG) via CRC 1375 NOA - Nonlinear Optics down to Atomic Scales – Project-ID 398816777, project A1.

%-----------------------------

%% Put the bibliography here, most people will use BiBTeX in
%% which case the environment below should be replaced with
%% the \bibliography{} command.

\begin{thebibliography}{30}

\bibitem{Kl-Rev}
D. N. Klyshko, Phys.-Usp. {\bf 39}, 573, (1996).

\bibitem{Atto1}
P. B. Corkum and F. Krausz, Attosecond science, Nat. Phys. {\bf 3}, 381 (2007).

%\bibitem{Chekhova} M.V. Chekhova, G. Leuchs, M. Żukowski, Bright squeezed vacuum: Entanglement of macroscopic light beams, Opt Commun, 337 (2015) 27-43.


%\bibitem{S1} A. Zavatta et al., Quantum-to-Classical Transition with Single-Photon-Added Coherent States of Light, Science {\bf 306}, 660-662 (2004).

%\bibitem{S2} A. Ourjoumtsev et al., Generating Optical Schrödinger Kittens for Quantum Information Processing, Science {\bf 312}, 83-86 (2006).

%\bibitem{S3} E.V. Mikheev et al., Efficient production of large-size optical Schrödinger cat states, Sci. Rep. {\bf 9}, 14301 (2019).

\bibitem{Gonoskov} I.A. Gonoskov, N. Tsatrafyllis, I.K. Kominis, P. Tzallas, Quantum optical signatures in strong-field laser physics: Infrared photon counting in high-order-harmonic generation, Sci Rep, 6 (2016) 32821.

\bibitem{Tzallas} N. Tsatrafyllis, I.K. Kominis, I.A. Gonoskov, P. Tzallas, High-order harmonics measured by the photon statistics of the infrared driving-field exciting the atomic medium, Nat Commun, 8 (2017) 15170.

\bibitem{Theidel} D. Theidel, V. Cotte, R. Sondenheimer, V. Shiriaeva, M. Froidevaux, V. Severin, A. Merdji-Larue, P. Mosel, S. Fröhlich, K.-A. Weber, U. Morgner, M. Kovacev, J. Biegert, H. Merdji, Evidence of the Quantum Optical Nature of High-Harmonic Generation, Prx Quantum, 5 (2024) 040319.

\bibitem{ParisBN}
P. Stammer, et al. \textit{High Photon Number Entangled States and Coherent State Superposition from the Extreme Ultraviolet to the Far Infrared},
Phys. Rev. Lett. 128, 123603, (2022).

\bibitem{Lewenstein} M. Lewenstein, M.F. Ciappina, E. Pisanty, J. Rivera-Dean, P. Stammer, T. Lamprou, P. Tzallas, Generation of optical Schrödinger cat states in intense laser–matter interactions, Nat Phys, 17 (2021) 1104-1108.



\bibitem{Rivera} J. Rivera-Dean, P. Stammer, A.S. Maxwell, T. Lamprou, A.F. Ordóñez, E. Pisanty, P. Tzallas, M. Lewenstein, M.F. Ciappina, Nonclassical states of light after high-harmonic generation in semiconductors: A Bloch-based perspective, Phys Rev B, 109 (2024) 035203.

\bibitem{Lamprou}
Th. Lamprou, P. Stammer, J. Rivera-Dean, N. Tsatrafyllis, M. F. Ciappina, M. Lewenstein, P. Tzallas, \textit{Recent developments in the generation of non-classical and entangled light states using intense laser-matter interactions}, arXiv:2410.17452, (2024).

\bibitem{Stammer_PRL} P. Stammer, J. Rivera-Dean, A.S. Maxwell, T. Lamprou, J. Argueello-Luengo, P. Tzallas, M.F. Ciappina, M. Lewenstein, Entanglement and Squeezing of the Optical Field Modes in High Harmonic Generation, Phys Rev Lett, 132 (2024) 143603.

\bibitem{Stammer0}
P. Stammer, J. Rivera-Dean, M. Lewenstein, \textit{Theory of quantum optics and optical coherence in high harmonic generation}, arXiv:2504.13287, (2025).

\bibitem{Stammer_PRX} P. Stammer, J. Rivera-Dean, A. Maxwell, T. Lamprou, A. Ordóñez, M.F. Ciappina, P. Tzallas, M. Lewenstein, Quantum Electrodynamics of Intense Laser-Matter Interactions: A Tool for Quantum State Engineering, Prx Quantum, 4 (2023) 010201.

\bibitem{Gonoskov_2021} I. Gonoskov, S. Gräfe, Light-matter quantum dynamics of complex laser-driven systems, J Chem Phys, 154 (2021) 234106.

\bibitem{Ivanov} S. Yi, N.D. Klimkin, G.G. Brown, O. Smirnova, S. Patchkovskii, I. Babushkin, M. Ivanov, Generation of Massively Entangled Bright States of Light during Harmonic Generation in Resonant Media, Phys Rev X, 15 (2025) 011023.

\bibitem{Dan}
C. S. Lange, T. Hansen, L. B. Madsen, \textit{Electron-correlation induced nonclassicallity of light from high-harmonic generation}, Phys. Rev. A 109, 033110, (2024).

\bibitem{Gorlach2} A. Gorlach et al, “The quantum-optical nature of high harmonic generation”, Nat. Comm. 11, 4598 (2020) 
\bibitem{Rubio-Neufeld} S. de la Peña et al. “Fully quantum theory of strong-field driven tunable entangled multi-photon states in HHG” arXiv:2512.03987v1 

\bibitem{Gonoskov_2024}
I. A. Gonoskov, et al., \textit{Nonclassical light generation and control from laser-driven semiconductor intraband excitations}, Phys. Rev. B, 109, 125110 (2024).

\bibitem{Tzur_2023} M. Even Tzur, M. Birk, A. Gorlach, M. Krüger, I. Kaminer, O. Cohen, Photon-statistics force in ultrafast electron dynamics, Nat Photonics, 17 (2023) 501-509.


\bibitem{Fang} Y. Fang, F.X. Sun, Q. He, Y. Liu, Strong-Field Ionization of Hydrogen Atoms with Quantum Light, Phys Rev Lett, 130 (2023) 253201.



\bibitem{Gorlach} A. Gorlach, M.E. Tzur, M. Birk, M. Krüger, N. Rivera, O. Cohen, I. Kaminer, High-harmonic generation driven by quantum light, Nat Phys, 19 (2023) 1689-1696.

\bibitem{Tzur_2024} M. Even Tzur, O. Cohen, Motion of charged particles in bright squeezed vacuum, Light Sci Appl, 13 (2024) 41.


\bibitem{EF1}
{Abedi, A. and Maitra, N. T. and Gross, E. K. U.},
\textit{Exact Factorization of the Time-Dependent Electron-Nuclear Wave Function},
Phys. Rev. Lett., {\bf 105}, 123002, (2010).





\bibitem{LargeNum}
L. Lachman, L. Slodička, and R. Filip, \textit{Nonclassical light from a large number of independent single-photon emitters}, Sci. Rep. 6, 19760 (2016).


%\bibitem{Dan2} Christian Saugbjerg Lange, Lars Bojer Madsen, \textit{Hierarchy of approximations for describing quantum light from high-harmonic generation: A Fermi-Hubbard model study}, Phys. Rev. A 111, 013113, (2025).

%\bibitem{Fleck}J. Fleck, J. Morris, and M. Feit, \textit{Time-Dependent Propagation of High Energy Laser Beams through the Atmosphere}, Applied Physics A: Materials Science and Processing, Vol. 10, No. 2, pp. 129-160, (1976).

\bibitem{Ville} 
V. J. Härkönen, and I. A. Gonoskov, \textit{On the diagonalization of quadratic Hamiltonians}, Journal of Physics A: Mathematical and Theoretical 55 (1), 015306 (2022).


\bibitem{my1q}
Gonoskov I.A., Vugalter G.A., Mironov V.A., \textit{Ionization in a Quantized Electromagnetic Field}, J. Exp. Theor. Phys. {\bf 105}, 1119, 5 (2007).


 

\end{thebibliography}
%\newpage

\section*{Methods/Appendix}

\subsection*{Derivation}
We start from the time-dependent Schr\"odinger equation for the fully quantum system in the Sch\"odinger representation within the dipole approximation for all light modes, see also \cite{Gonoskov_2021} and references therein. Starting point is the Pauli-Fierz Hamiltonian, in eq.~\ref{eq:Pauli_Fierz}. As outlined above, for the primary derivation, we restrict ourselves to one mode of emitted light ($\Omega$, $q$, $\beta$, see \cite{Gonoskov_2021}) and an intense classically-described driving laser pulse with the vector potential $\vec{A}_{class}(t)$ polarized along $x$-axis. Within the commonly evoked single-active electron approximation, we transform into interaction picture using $U_{I}=\exp[-\mathrm{i}\Omega{}t\hat{N}]$ and subsequently apply a quantum optical analog of the gauge transformation, $\exp\left\{\mathrm{i}{}x\,\frac{e}{\hbar c}\left[{A}_{c}(t)+\hat{A}_{\Omega}(t)\right]\right\}$ to obtain:

\begin{equation}\label{n1m}
\mathrm{i}\hbar \dfrac{\partial \Psi }{\partial t}=\bigg[ \hat{H}_{0}+e\,x\cdot{}\left[ F_{c}\left( t\right) +\hat{F}_{\Omega}\left( t\right) \right] \bigg] \Psi \,,
\end{equation}
where $\hat{H}_{0}$ is the initial unperturbed Hamiltonian, $F_{class}\left( t\right)=-(1/c)\,\dot{A}_{c}(t)$ is the force of the classically-described driving laser pulse, and the force operator of the $\Omega$-mode $\hat{F}_{\Omega}\left( t\right)=\dfrac{\Omega }{c}\beta q\sin \left( \Omega t\right) +\mathrm{i}\beta\dfrac{\Omega }{c}\dfrac{\partial }{\partial q}\cos \left( \Omega t\right) $ (eq.~\ref{eq:force_operator}).\\

A useful approximation derived from the exact factorization \cite{EF1}, leading to a set of equations of reduced dimensionality, is the parametric factorization, see \cite{Gonoskov_2021}. For the considered system, this gives (including partial normalization conditions and up to a certain factorization gauge):

\begin{equation}\label{n4m}
\Psi\left( \vec{r},\,\beta{}, q\,,t\right) =\psi^{el} \left( \vec{r},\,\beta{}q\,,t\right) \cdot \phi^{light}\left( q,\beta{},t\right) + \mathcal{O}(\beta)\,.
\end{equation}

Direct substitution gives:

\begin{equation}\label{n5m}
\mathrm{i}\hbar\,\dfrac{\partial \psi^{el} }{\partial t}\phi^{light}+\mathrm{i} \hbar \,\psi^{el} \dfrac{\partial \phi^{light}}{\partial t}=\phi^{light}\hat{H}_{0}\psi^{el} +e\,x\cdot \left[ F_{class}\left( t\right) +\hat{F}_{\Omega}\left( t\right) \right]\psi^{el} \phi^{light}\,.
\end{equation}
An advantage of parametric factorization for this equation is that \bl{$\hat{F}_{\Omega}\left( t\right)\psi^{el} \phi^{light}\approx{}\psi^{el} \cdot{} \hat{F}_{\Omega}\left( t\right)\phi^{light}$ (up to higher-order perturbations) and, thus,} $|\,\langle \beta\hat{p}_{\Omega} \rangle _{\psi^{el}}\,|\ll{}|\,\langle \beta\hat{q}_{\Omega} \rangle _{\psi^{el}}\,|$ \bl{(by one till three orders of magnitude in or case)}, so the corresponding non-local operator can be neglected. Then we can rewrite the equation:
\begin{equation}\label{n5g1}
\phi^{light}\left[\mathrm{i} \hbar\dfrac{\partial }{\partial t}-\hat{H}_{0}\right]\psi^{el}+\mathrm{i}\hbar \,\psi^{el} \dfrac{\partial \phi^{light}}{\partial t}= +e\,x\,\psi^{el} \cdot \left[ F_{class}\left( t\right) +\hat{F}_{\Omega}\left( t\right) \right] \phi^{light}\,.
\end{equation}

Since the parametric factorization Eq.(\ref{n4m}) can be chosen with an arbitrary function $\chi(\beta{}q,t)$ meeting the conditions of partial normalization $\psi^{el} \left( \vec{r},\,\beta{}q\,,t\right)\,\exp({-\mathrm{i}\chi(\beta{}q,t)}) \cdot \phi^{light}\left( q,\beta{},t\right)\,\exp({+\mathrm{i}\chi(\beta{}q,t)})$, a specific $\chi(\beta{}q,t)$ has to be chosen for further considerations \cite{Gonoskov_2021}. This choice is usually called a gauge of factorization. Here, we present a {special gauge of the parametric factorization} by choosing the function $\chi(\beta{}q,t)$ in the following way:
\begin{equation}\label{n5g2}
\chi(\beta{}q,t)\;:\;\;\;\int\limits_{-\infty}^{+\infty}\psi^{el,*} \left( \vec{r},\,\beta{}q\,,t\right)\,\left[\mathrm{i}\hbar\dfrac{\partial }{\partial t}-\hat{H}_{0}\right]\,\psi^{el} \left( \vec{r},\,\beta{}q\,,t\right)\,d^{3}r\,=0\,.
\end{equation}

With this condition, we multiply both sides of Eq.(\ref{n5g1}) by $\psi^{el,*} \left( \vec{r},\,\beta{}q\,,t\right)$ and integrate over $d^{3}r$. Since, according to the partial normalization condition \cite{Gonoskov_2021} $\int\limits_{-\infty}^{+\infty}|\psi^{el} \left( \vec{r},\,\beta{}q\,,t\right)|^{2}\,d^{3}r=1$, we obtain eq.~\ref{eq:TDSE-q}:
\begin{equation}\label{n5g3}
\mathrm{i}\hbar\,\dfrac{\partial \phi^{light}}{\partial t} =  +e\,\langle x\rangle _{\psi^{el}}\cdot \left[F_{class}\left( t\right) + \hat{F}_{\Omega}\left( t\right)\right] \phi^{light} \,.
\end{equation}

The solution of this equation -- in case  the dipole moment $\langle x\rangle _{\psi^{el}}$ depends only on time (with a coherent or vacuum initial state) -- is a coherent state (with a Gaussian shape as a function of $q)$. The $\beta{}q$-dependence in $\langle x\rangle _{\psi^{el}}$ may lead to the nonclassical solutions with non-Gaussian shapes.

%This equaton looks similar to Eq.(\ref{n2}) but has a significant difference: the dipole moment $\langle x\rangle _{P\Phi}$ depends parametrically on $\beta{}q$ thus, the possible solutions of Eq.(\ref{n5g3}) are not so trivial like Eq.(\ref{n3}) in case of basic factorization. 
To obtain the dipole moment $\langle x\rangle _{\psi^{el}}=\int\limits_{-\infty}^{+\infty}x\cdot{}|\psi^{el} \left( \vec{r},\,\beta{}q\,,t\right)|^{2}\,d^{3}r$, we need to find an explicit expression for $\psi^{el} \left( \vec{r},\,\beta{}q\,,t\right)$ up to an arbitrary exponential phase factor via Eqs.(\ref{n5m}-\ref{n5g1}). 

Let us assume that in some time interval $\phi^{light}\left( q,\beta{},t\right)\neq{}0$ (the converse should be considered separately by direct analysis of Eq.(\ref{n5g1}) and related statements). This is true, given some moderate changes of initial coherent or vacuum state $\phi^{light}_{0}\left( q\right)=C\,\exp\left[ -\left( q-q_{0} \right) ^{2}/2\right]\neq{}0$. In this case, we can multiply both sides of Eq.(\ref{n5g1}) by $\left(\phi^{light}( q,\beta{},t)\right)^{-1}$ and obtain:
\begin{equation}\label{n5g4}
\mathrm{i} \hbar \dfrac{\partial \psi^{el} }{\partial t}+\mathrm{i} \hbar \left(\phi^{light}\right)^{-1} \dfrac{\partial \phi^{light}}{\partial t}\cdot \psi^{el} =\hat{H}_{0}\psi^{el} +e\,x \,\psi^{el}\cdot \left[ F_{class}\left( t\right) + \left(\phi^{light}\right)^{-1} \hat{F}_{\Omega}\left( t\right) \phi^{light} \right] \,.
\end{equation}

Since $\mathrm{i}\hbar \left(\phi^{light}\right)^{-1} \dfrac{\partial \phi^{light}}{\partial t}\cdot \psi^{el}$ corresponds to a phase factor and has no effect to $\langle x\rangle _{\psi^{el}}$, we transform:
\begin{equation}\label{n5g5}
\mathrm{i}\hbar\, \dfrac{\partial \psi^{el} }{\partial t}=\hat{H}_{0}\psi^{el} + e\,x\cdot\bigg[ F_{class}\left( t\right) + F_{\phi^{light}}\left( \beta{}q\,,t\right) \bigg]\,\psi^{el} \,,
\end{equation}
where $F_{\phi^{light}}\left( \beta{}q\,,t\right)=\,\left(\phi^{light}\right)^{-1} \hat{F}_{\Omega}\left( t\right) \phi^{light} $ is the parametric representation of the force operator (eq.~\ref{eq:force_operator}), which is for nearly coherent states of moderate to high intensities $\,\approx\, \dfrac{\Omega }{c}\beta q\sin \left( \Omega t\right) $ such that $q \approx \langle q\rangle$. We now can write a complete system of equations describing the evolution of the quantum state $\phi^{light} \left( q,\beta{},t\right)$ of a given light mode in a special parametric factorization description:

\begin{equation}\label{i4}
\begin{aligned}
&\mathrm{i} \hbar\,\dfrac{\partial \phi^{light}}{\partial t} =  +e\,\langle x\rangle _{\psi^{el}}\cdot \left[F_{class}\left( t\right) + \hat{F}_{\Omega}\left( t\right)\right] \phi^{light} \,,\\
&\mathrm{i}\hbar\, \dfrac{\partial \psi^{el} }{\partial t}=\hat{H}_{0}\psi^{el} + e\,x\cdot\bigg[ F_{class}\left( t\right) + F_{\phi^{light}}\left( \beta{}q\,,t\right) \bigg]\,\psi^{el} \,,\\
&\langle x\rangle _{\psi^{el}}=\,\int\limits_{-\infty}^{+\infty}x\cdot{}|\psi^{el} \left( \vec{r},\,\beta{}q\,,t\right)|^{2}\,d^{3}r\,,\;\;\;\;F_{\phi^{light}}\left( \beta{}q\,,t\right)=\,\left(\phi^{light}\right)^{-1} \hat{F}_{\Omega}\left( t\right) \phi^{light} \,.
\end{aligned}
\end{equation}

It can be seen that the nonlinear dependence on $q$ arises from the additional $q$ dependence of the dipole moment $\langle x\rangle _{\psi^{el}}$, resulting from the $q$-dependent, parametrically connected electronic state $\psi^{el}=\psi^{el} \left( \vec{r},\,\beta{}q\,,t\right)$. Thus, the nonlinear terms in equation Eq.(\ref{n5g3}) can lead to the generation of non-classical states of light, even if the initial state $\phi^{light}_{0}\left( q\right)$ is a normal vacuum or a coherent state.

\subsection*{Numerical Scheme}

Fig.~\ref{fig:setup} displays the additional information on the systems described and investigated above: the 1D atomic Ca-atom model and the 1D CaO molecular model system interacting with an intense classical laser pulse. \bl{As commonly done in strong-field physics, a soft-core Coulomb potential model was used ($\sim{}1/\sqrt{(x-x_{n})^{2}+d^{2}}\;$). We chose the parameter $d^{2}$ so that the model system corresponds to realistic ground state energies, for example (in a.u.): $2.002$ for hydrogen model; $12.448$ for Ca-atom model; $21.264$ and $25.0$ for CaO molecular model}. The initial states are either the system's corresponding electronic ground states (orange shaded) or superposition states (green shaded). The driven atom/molecular systems generate harmonic light $\Omega$, the time dependence of which is displayed in on the right-hand side.

\begin{figure}
	\includegraphics[width=\columnwidth]{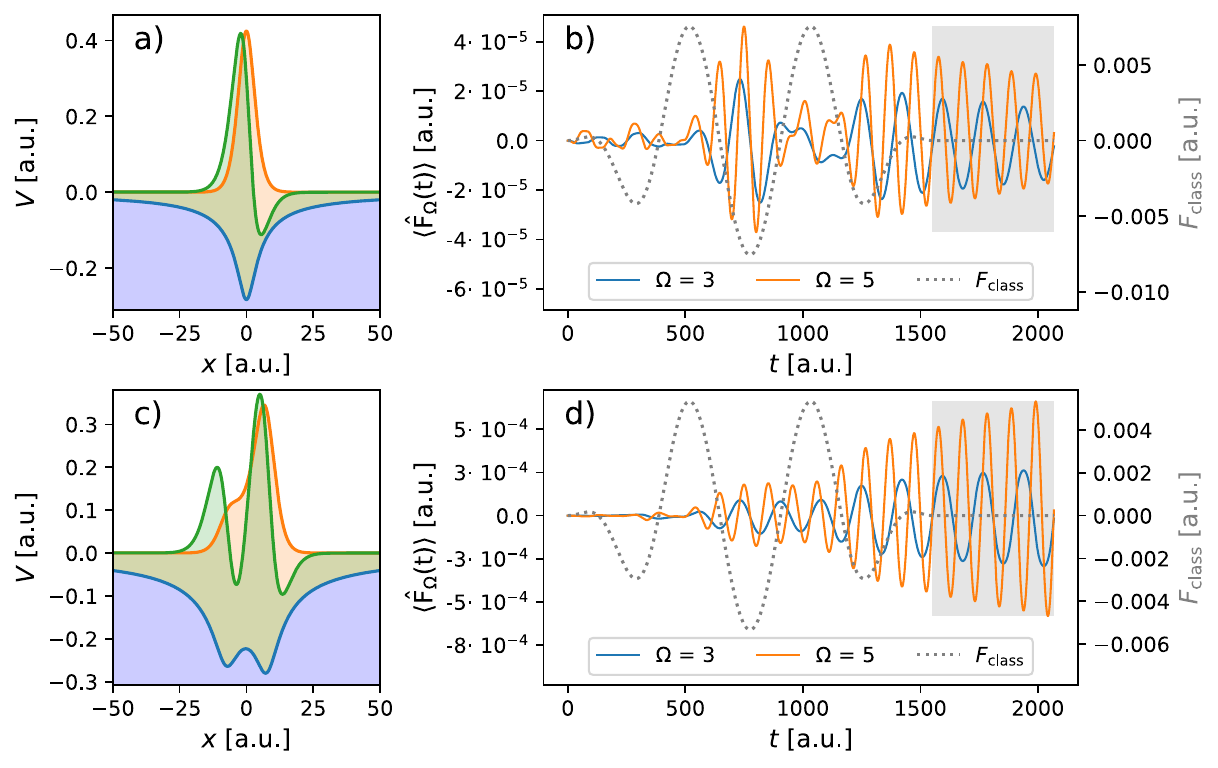}
	\caption{{Nonclassical harmonic generation from various model systems discussed in the main text: 1D electronic system coupled to one mode of quantized light with harmonic frequency $\Omega$ as labeled. (A) {Left}: 1D soft-core model potential (blue) mimicking the Ca-atom and corresponding electron wavefunctions of the ground state (orange) and a superposition of the ground and excited electronic state  (Ca atom: I$_{p}$ = 6.11 eV). {Right}: Solution of the electronic Schr\"odinger equation (eq.~\ref{eq:TDSE-x}), displaying the time-dependent electron dipole moment (eq.~\ref{eq:dipole}). The results of two calculations are shown, where the field-driven electronic Schr\"odinger equation is coupled to a light mode of $\Omega=3$ and $\Omega=5$. For comparison, the dashed line displays the (classical) driving laser, with parameters $F_{class}(t)\;$ ($\lambda$ = 3750nnm, T$_{p}$ = 2 cycles), $I = 2\cdot 10^{12}$ W/cm$^{2}$ $\beta = 0.55 \text{au.}$.
			(B) same for an asymmetric diatomic molecule model (1D CaO model with I$_{p}$ = 6.6 eV), however with $\beta$ = 0.82 au., $I = 10^{12}$ W/cm$^{2}$. The gray area shows the modes' behavior after the interaction with the driving laser.}}\label{fig:setup}
\end{figure}

The procedure is general, allowing the investigation of a large number of emitters $N_e$ to be arbitrary quantum systems with various initial states. The numerical solution scheme for the multi-emitter case is explained in what follows.

We consider an ensemble of Coulomb-isolated quantum emitters in a resonator, driven by a classical laser pulse of frequency $\omega$ and one single quantized light mode of frequency $\Omega$. Each emitter (index $n$) contributes with its oscillating dipole moment ($ d_n (\beta q, t)  = e_n\langle x(\beta q, t)\rangle_n$), generating radiation of frequency $\Omega$. The collective response
\begin{equation}
	D(\beta q,t)=\sum_{n=1}^{N_e} d_n(\beta q, t)
\end{equation}
can, under suitable conditions, generate intense non-classical radiation due to the large number of emitters $N_e$, as shown in Fig.~\ref{fig:intense} and Fig.~\ref{fig:overview_mulitemitter}.

To describe the evolution of the quantized mode, represented by the light-state wavefunction ($\phi^{\text{light}}(q,\beta,t)$), we employ a transformation-based solution strategy \cite{my1q,Ville}. The interaction term in the light-mode equation contains a nonlocal operator proportional to
\begin{equation}
	i\langle x\rangle_{\psi^{el}}(q,t)\,\cos(\Omega t)\,\frac{\partial}{\partial q},
\end{equation}
which induces both a shift and a deformation of the $q$-coordinate. For sufficiently small $\beta q$, the dipole response can be expanded as
\begin{equation}
	\langle x\rangle_{\psi^{el}}(\beta q,t)
	\approx
	\left.\langle x\rangle_{\psi^{el}}(t)\right|_{F_{class}(t)}+ (\beta q)f_{1}(t) + (\beta q)^2 f_{2}(t)+\cdots ,
\end{equation}
with $f_{1}(t)\sim \cos(\Omega t)\,\left.\partial\langle x\rangle_{\psi^{el}}/\partial F\right|_{F=F_{class}(t)} $ etc. This structure naturally leads to a hierarchy of generalized shift operators,
\begin{equation}
	\exp\left[a_0(t)\frac{\partial}{\partial q}\right],\qquad
	\exp\left[a_1(t)\,q\,\frac{\partial}{\partial q}\right],\qquad
	\exp\left[a_2(t)\,q^2\,\frac{\partial}{\partial q}\right],\ \dots,
\end{equation}
closely related to extended forms of the transformations introduced in \cite{my1q,Ville}. These operators collectively generate the full $q$-dependence of the light-state evolution.

\begin{figure}
	\includegraphics[width=\columnwidth]{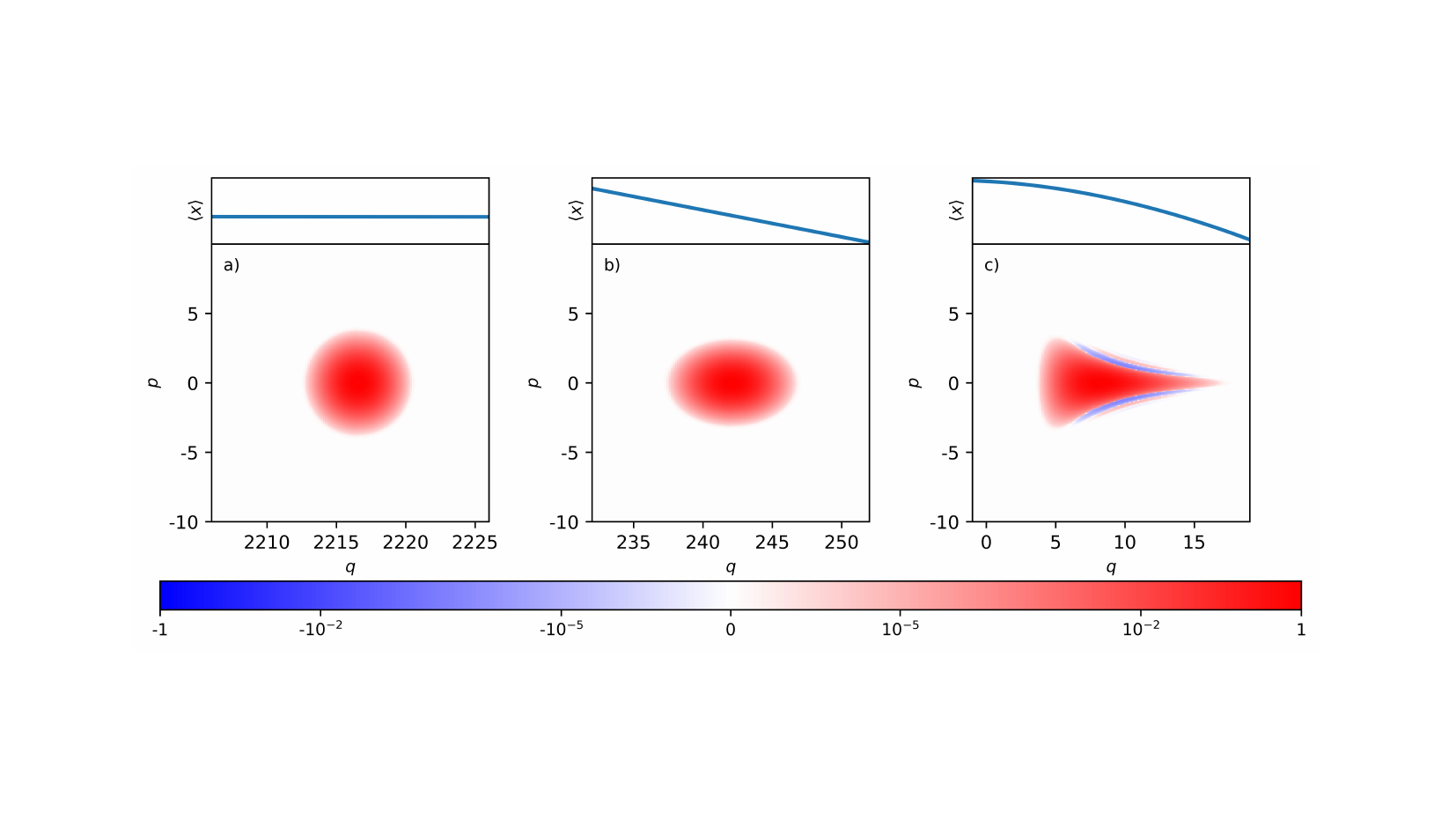}
	\caption{As in Figure \ref{fig:overview}, visualization of the emergence of nonclassicality in high-order harmonic generation to result from the nonlinear $q$- dependence of the oscillating dipole moment, however for the multi-emitter case, demonstrating that high photon numbers can be obtained. The upper row displays the dipole moments expanded into a series (truncated after the terms as labeled) (eq.~\ref{eq:dipole-series}) while the lower row shows the resulting output state of light in phase space, represented by its Wigner function. (a) Constant, almost $q$-independent dipole moment leads to (below): emitted light represented by a Wigner function of an almost coherent state with high photon number (here $\Omega=5$th harmonic). (b) Linear $q$-dependence of dipole moment can lead to Wigner function of a squeezed state with high photon number ($\Omega=5$th harmonic). (c) Dipole moment containing terms nonlinear in $q$, here for an approximately quadratic dependence on $q$, can lead to the generation of nonclassical light, as represented by a Wigner function with negative values/regions ($\Omega=5$th harmonic).  Initial conditions: product state of light and matter, with the initial system being in the second excited state for the H-atom; Laser parameters: (a) and (b): $F_{class}(t)\;$ ($\lambda$ = 2227 nm, T$_{p}$ = 2 cycles,  $I = 10^{13}$ W/cm$^{2}$). The quantization parameter is $\beta = 1.4\cdot 10^{-6} \text{au.}$ in (a) and $\beta = 1.4\cdot 10^{-3} \text{au.}$ in (B). The corresponding parameters for (c) are $F_{class}(t)\;$ ($\lambda$ = 3000 nm, T$_{p}$ = 2 cycles,  $I = 2\cdot 10^{12}$ W/cm$^{2}$) and $\beta = 0.22 \text{au.}$ The average number of photons in the output state is approximately $N_{\Omega}\approx{\bar{q}^{2}/2}$}\label{fig:overview_mulitemitter}
\end{figure}

In practice, the dominant long-time contribution arises from the resonant harmonic at frequency $\Omega$, whose effect accumulates linearly with the interaction time. Assuming that the electronic response is identical for each laser pulse, and that the resonant harmonic dominates the non-classical buildup, we replace the explicit time dependence by the real part of the Fourier coefficient $ d_{\Omega}(q)$. The evolution operator for the light mode then takes the form
\begin{equation}
	\exp\left[-\frac{\beta\Omega}{c}\,d_{\Omega}(q) \, t_{\mathrm{int}} \,\partial_q \right]
	\phi^{\text{light}}_{0}(q) = \exp\left[a\,f_{\Omega}(q)\,\partial_q\right] \,\phi^{\text{light}}_{0}(q),
\end{equation}
where ($t_{\mathrm{int}}$) is the interaction time and $f_{\Omega}(q)$ is an appropriate polynomial or rational fit to $d_\Omega(q)$, obtained from the TDSE simulations of the electronic subsystem (Fig.~\ref{Fig5}). In contrast to the simple exemplary cases in the main text, Fig.~\ref{fig:overview} and Fig.~\ref{fig:overview_mulitemitter}, where a the dipole moment was expanded into the series and truncated after specific terms, Fig.~\ref{Fig5} displays now the results as obtained from the electronic TDSE, however, an average over the interaction time, rather than a specific time, i.e.

\begin{equation}\label{eq:dipole_omega}
	\langle x\rangle_\Omega \propto \int^{t_{int}} \langle x(\Omega \beta q/c, t)\rangle \cos \left( \Omega t\right)\,dt .
\end{equation}

In the simplest multi-emitter case, this dipole scales as ($N_e$), which strongly favors the analytical approach because direct numerical integration would involve rapidly growing operator amplitudes.

\begin{figure}
	\includegraphics[width=0.5\columnwidth]{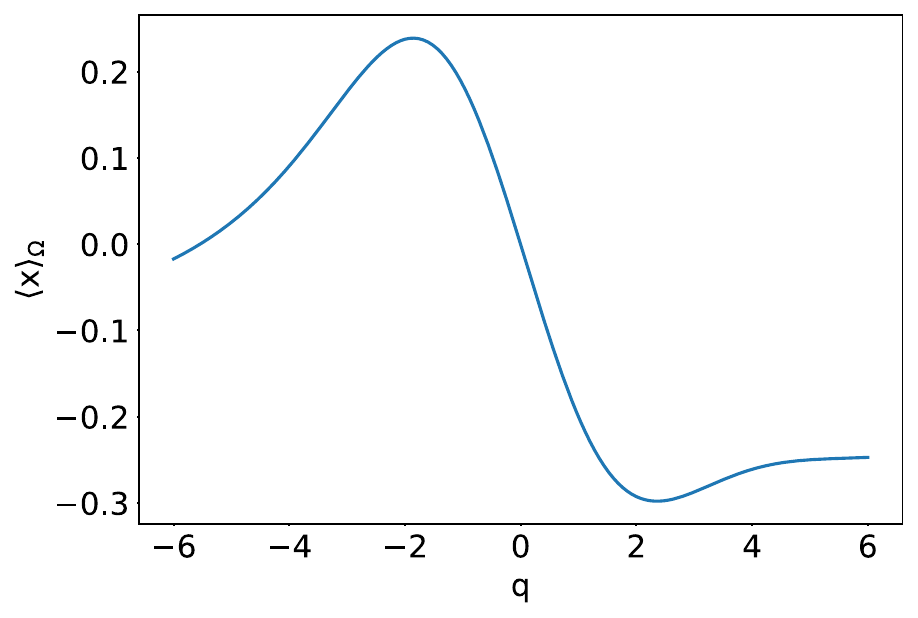} 
	\caption{Resonant part of dipole moment (eq.~\ref{eq:dipole_omega}) as a function of $q$, corresponding to the numerical solutions in a field $F_{class}(t)+y\cdot{}\sin(\Omega{t})$; as described in the text, polynomial fitting this form is performed, which is used for the analytical solving the $\phi^{light}(q,\beta,t_{\text{int}})$. The calculations were for a model hydrogen atom with the following parameters: $\Omega=13$, $\lambda$ = 2227 nm, T$_{p}$ = 4 cycles, $I$ = 4 $\cdot$ 10$^{13}$ W/cm$^{2}$.}\label{Fig5}
	\end{figure}

Choosing $f_\Omega(q)$ such that the shift has a closed-form real action on $\phi^{\text{light}}_0(q)$ yields
\begin{equation}
	\exp\left[a,f_{\Omega}(q)\partial_q\right]\,\phi^{\text{light}}_0(q) = \phi^{\text{light}}_0 \big(Q^{-1}[Q(q)+a]\big),
	\quad Q(q) =\int_0^q\frac{dq'}{f_\Omega(q')}.
\end{equation}
Therein, $Q(q)$ is a polynomial series in $q$, with coefficients depending on the expansion of the (resonant part of the) dipole moment, and $Q^{-1}$ is the inverse procedure, such that $Q^{-1}[ Q(q)] = q$. This procedure can be calculated exactly, and, particularly for small term expansion series, also very efficiently. This analytical expression captures both the $q$-shift and the nonlinear deformation induced by the emitters’ dipole response. Examples in Fig.~\ref{fig:intense}  demonstrate that multi-emitter systems with strong excitation resonances produce pronounced non-classical features, which can be further enhanced by choosing an initially non-classical light state.

We emphasize that the scheme presented is universally applicable for atoms, molecules but also semiconductors: The procedure generalizes directly to strongly driven semiconductors by summing the single-emitter dipoles to obtain effective intra- and inter-band currents. The resulting non-classical emission follows analogously and is fully compatible with intraband models such as \cite{Gonoskov_2024}.

\end{document}